# Mathematical modeling of shear-activated targeted nanoparticle drug delivery for the treatment of aortic diseases

*Yonghui Qiao, Yan Wang, Yanlu Chen, Kun Luo\* and Jianren Fan\**

*State Key Laboratory of Clean Energy Utilization, Zhejiang University, Hangzhou, China*

\* Corresponding author:

**Kun Luo**

State Key Laboratory of Clean Energy Utilization, Zhejiang University

38 Zheda Road, Hangzhou, China, 310027

Phone: +86-18667160183   E-mail：zjulk@zju.edu.cn

**Jianren Fan**

State Key Laboratory of Clean Energy Utilization, Zhejiang University

38 Zheda Road, Hangzhou, China, 310027

Phone: +86-13336107178   E-mail：fanjr@zju.edu.cn

**Acknowledgments**

This research was supported by the National Postdoctoral Program for Innovative Talents (CN) [grant number BX20200290], Postdoctoral Science Foundation (CN) [grant number 2020M681852], Postdoctoral Science Foundation of Zhejiang Province (CN) [grant number ZJ2020153].

**Conflict of interest**

All authors declare that they have no conflicts of interest.

**Abstract:** The human aorta is a high-risk area for vascular diseases, which are commonly restored by thoracic endovascular aortic repair. In this paper, we report a promising shear-activated targeted nanoparticle drug delivery strategy to assist in the treatment of coarctation of the aorta and aortic aneurysm. Idealized three-dimensional geometric models of coarctation of the aorta and aortic aneurysm are designed, respectively. The unique hemodynamic environment of the diseased aorta is used to improve nanoparticle drug delivery. Micro-carriers with nanoparticle drugs would be targeting activated to release nanoparticle drugs by local abnormal shear stress rate (SSR). Coarctation of the aorta provides a high SSR hemodynamic environment, while the aortic aneurysm is exposed to low SSR. We propose a method to calculate the SSR thresholds for the diseased aorta. Results show that the upstream near-wall area of the diseased location is an ideal injection location for the micro-carriers, which could be activated by the abnormal SSR. Released nanoparticle drugs would be successfully targeted delivered to the aortic diseased wall. Besides, the high diffusivity of the micro-carriers and nanoparticle drugs has a significant impact on the surface drug concentrations of the diseased aortic walls, especially for aortic aneurysms. This study preliminary demonstrates the feasibility of shear-activated targeted nanoparticle drug delivery in the treatment of aortic diseases and provides a theoretical basis for developing the drug delivery system and novel therapy.

# 1. Introduction

The predominant treatment for aortic diseases has transformed from traditional open surgery to thoracic endovascular aortic repair. However, postoperative complications associated with the implanted stent graft, such as endoleaks and retrograde type A dissection are still confusing clinicians (Czerny et al. 2021). Therefore, there is a pressing need for proposing new treatment modalities to avoid the risks caused by the endograft.

Targeted nanoparticle drug delivery may be a promising alternative with high efficiency and low side effects once the specific drugs for the treatment of aortic wall lesions are successfully developed. Nano-drugs are more and more prevalent due to their superior dissolution and absorption capacity (Pala et al. 2020). Targeted delivery could be performed by physical and chemical stimulus including temperature, magnetic field, and special chemical markers, etc. Recently, a novel biomimetic strategy, shear-activated targeted drug delivery was proposed to treat the narrowing of arterial blood vessels (Holme et al. 2012; Korin et al. 2012). Specifically, Nano-drugs would be released by nanoparticle aggregates or artificial lentil-shaped liposomes, when those nanocontainers expose to local high shear stress in stenotic blood vessels (Epshtein and Korin 2017).

Animal experiments were carried out to test the strategy of targeted nanoparticle drug delivery. Korin et al. (2012) reported the high efficacy of shear-activated nanoparticle aggregates to normalize the pulmonary artery pressure in a mouse pulmonary embolism model. Marosfoi et al. (2015) coupled shear-activated nanoparticle aggregates with temporary endovascular bypass to restore a rabbit common carotid artery occlusion. The superiority of shear-activated nanoparticle aggregates compared to other treatment methods was demonstrated. On the one hand, the total amount of drugs required is reduced and the nanoparticles could be targeted delivered to the vascular lesion. On the other hand, this strategy also reduces the nanoparticles reaching downstream of the vascular lesion. A small number of nanoparticles and inactivated nanoparticle aggregates mean a decreased risk of drug delivery to the healthy sites.

Image-based computational fluid dynamics is a powerful tool to study shear-activated targeted drug delivery and reveal the unique hemodynamic environment of diseased arteries. Ebrahimi et al. (2021) explored the performance of targeted Nano-drug delivery in abdominal aortic aneurysms and they found that liposomes are better than solid particles from the perspective of targeting efficiency to the inner wall. Meschi et al. (2021) applied continuum transport models to predict the release process of Nano-drug in coronary artery disease and the controlling role of Lagrangian coherent structures was evaluated. Nevertheless, previous work in the field of shear-activated targeted drug delivery only focuses on vascular stenosis and small blood vessels, the feasibility for aneurysms and large blood vessels has not received enough attention.

Stenosis and aneurysm have opposite shear-activated targeted drug delivery mechanisms. Stenosis vessels always accompany abnormal hemodynamics and the high shear stress rate (SSR) characteristic would be used to activate micro-carriers in the diseased region. Different from the vascular stenosis scenario, the aneurysm would create a low SSR environment. Nanoparticle drugs can be encapsulated in the high SSR micro-carriers. When exposed to a low SSR aneurysm, the micro-carriers collapse, and nanoparticles are released to the diseased vessel wall.

This primary study focuses on the shear-activated targeted drug delivery using the combination of micro-carriers and nanoparticles in aortic diseases, which may be a promising treatment in the future clinic. Firstly, we designed an idealized three-dimensional aortic geometry, where coarctation of the aorta and aortic aneurysm were constructed. Then the corresponding mathematical model was developed to investigate the feasibility of this new strategy of targeted drug delivery for treating aortic diseases. Besides, we proposed novel methods to inject the micro-carriers and calculate the SSR thresholds for coarctation of the aorta and aortic aneurysm. Finally, the crucial hemodynamics and distribution of micro-carriers and nanoparticles were quantitatively explored to elucidate the strength and limitation of shear-activated targeted drug delivery in the coarctation of the aorta and aortic aneurysm. We also revealed the impact of the high diffusivity of the micro-carriers and nanoparticle drugs on the surface drug concentrations of the diseased aortic walls.

## 2. Methodologies

### *2.1 Geometry construction and mesh generation*

Based on our previous study (Qiao et al. 2019a), an idealized aortic geometric model was constructed in SolidWorks (SolidWorks, Waltham, MA). We retained three supra-aortic branches while other relatively small branches of descending aorta were excluded for simplicity. Artificial coarctation and aneurysm were designed in the descending aorta to simulate two kinds of common aortic diseases (Fig. 1). The degree of coarctation is 75%, which is defined as the ratio of the reduced radius of the coarctation to the radius of descending aorta. The syringe indicates where the micro-carriers are injected from the near-wall area (1mm from the aortic wall). Specifically, coarctation of the aorta has an injection point, while the aneurysm has three.

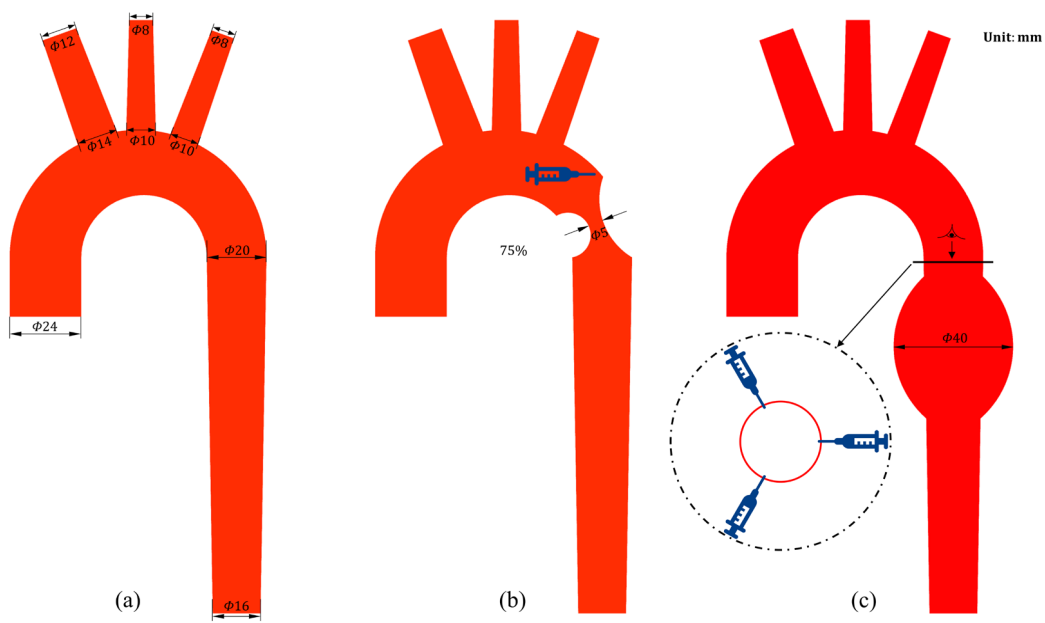

**Fig. 1** Schematic diagram of the idealized aortic geometric models. (a) Healthy aorta. (b) Coarctation of the aorta. The degree of coarctation is 75%, which is defined as the difference between one and the radius ratio of coarctation to the descending aorta. (c) Aortic aneurysm. The syringe indicates where the micro-carriers are released from the near-wall area (1mm from the wall).

The blood flow domains were divided into unstructured meshes, which consisted of tetrahedral elements in the flow center region and eight prism layers near the wall. It should be emphasized that the prism layers are essential to acquire accurate SSR distribution on the aortic wall. The maximal element size was 3E-4 m and the height of the first layer was set to 1.5E-5 m to satisfy the requirements of the turbulent model (y+ < 1). The final fluid domains had more than 5,700,000 elements. Grid-independence tests were carried out and the difference in wall shear stress was less than 2%. Therefore, the grid numbers of the computational domains are reasonable.

*2.2 Numerical model and computational details*

The blood flow was considered as incompressible and non-Newtonian. We applied the Carreau-Yasuda viscosity model to capture the blood shear thinning characteristics (Gijsen et al. 1999). The aortic wall was assumed to be rigid with no slip. A physiological pulsatile blood flow waveform (Alastruey et al. 2016) was coupled in the inlet of the ascending aorta and we used a parabolic distribution to account for the development of blood flow injected from the left heart.

$$u(r)=2u_{ave}(1-\frac{r^2}{R_{inlet}^2}) \tag{1}$$

where $u_{ave}$ is the inlet averaged velocity, $r$ is the radial location and R is the inlet radius. The three-element Windkessel model was adopted to calculate the pulsating pressure waveform in the four aortic outlets. The workflow proposed by Pirola et al. (2017) was applied to determine the three parameters of the Windkessel model. Specifically, the healthy upper body was assumed to receive a 30% blood flow rate, which was distributed according to the respective outlet area of three supra-aortic branches. The systolic and diastolic pressures were 120 and 80 mmHg, respectively. Table 1 shows the parameters of the Windkessel model and the same set of model parameters were applied in the other two diseased scenarios.

Turbulence should be considered in the coarctation region, which has been assessed by clinical experimental measurement (Arzani et al. 2012; Lantz et al. 2013). The turbulent kinetic energy in aortic coarctation could be acquired using a high-resolution large-eddy simulation or direct numerical simulation method (Arzani et al. 2012; Lantz et al. 2013). Considering the limited computational resource, we chose the shear stress transport (SST) model to predict the turbulent blood flow, which has been validated in stenosis and aneurysm (Tan et al. 2008; Tan et al. 2009). At the ascending aorta inlet, the turbulence intensity was specified to be 1.5% (Tan et al. 2009).

Table 1. Parameters of the three-element Windkessel model.

| OUTLET | $R_1$ [$10^7$ Pa s m$^{-3}$] | $C$ [$10^{-10}$ m$^3$ Pa$^{-1}$] | $R_2$ [$10^8$ Pa s m$^{-3}$] |
|---|---|---|---|
| BT | 6.058 | 21.76 | 7.621 |
| LCA | 15.49 | 9.620 | 17.06 |
| LSA | 15.49 | 9.620 | 17.06 |
| DA | 3.118 | 95.66 | 1.559 |

$R_1$: proximal resistance; $R_2$: distal resistance; $C$: vessel compliance; BT: brachiocephalic artery; LCA: left carotid artery; LSA: left subclavian artery; DA: descending aorta.

All the simulations were performed on ANSYS Workbench (ANSYS Inc, Canonsburg, USA). We adopted a constant time step of 1 ms and the fifth cardiac cycle data were post-processed to present periodic hemodynamic results.

### 2.3 Targeted drug delivery using shear-sensitive micro-carriers

Micro-carriers are micron-sized aggregates of drug-coated nanoparticle drugs, which could be passively transported with the pulsating blood flow. When the local SSR is higher or lower than a certain threshold, micro-carriers would collapse and release nanoparticle drugs. The concentrations of the micro-carriers and nanoparticles are controlled by a coupled system of continuum advection–diffusion-reaction equations (Meschi et al. 2021).

$$\frac{\delta C_m}{\delta t} + u \cdot \nabla C_m = \nabla \cdot (D_m \cdot \nabla C_m) - k_c C_m \tag{2}$$

$$\frac{\delta C_n}{\delta t} + u \cdot \nabla C_n = \nabla \cdot (D_n \cdot \nabla C_n) + k_c C_m \tag{3}$$

where m and n are micro-carriers and nanoparticle drugs, respectively, $D$ indicates the diffusion coefficient, which consists of the contributions of Brownian motion and shear-induced diffusion (Grief and Richardson 2005).

$$D = D_{Br} + D_{sh} \tag{4}$$

$$D_{Br} = kT/(6\pi\mu_p r) \tag{5}$$

$$D_{sh} = K_{sh}\dot{\gamma}r_{RBC}^2 \tag{6}$$

where $k$ is Boltzmann's constant, $T$ represents the absolute temperature, $\mu_p$ is plasma viscosity, $r$ indicates the corresponding particle radius, $\dot{\gamma}$ is SSR, and $r_{RBC}$ is the radius of red blood cells. It should be noted that $D_{sh}$ is not related to the particle dimensions and the value for micro-carriers and nanoparticles are the same. There is a coefficient ($k_c$) in the reaction source term and its value is controlled by the collapsing SSR threshold. Coarctation of the aorta is characterized by local high SSR, where micro-carriers would release nanoparticle drugs.

$$k_c = \begin{cases} k_{col} & \text{if } \dot{\gamma} \geq \dot{\gamma}_{threshold} \\ 0 & \text{if } \dot{\gamma} < \dot{\gamma}_{threshold} \end{cases} \tag{7}$$

While the aortic aneurysm provides a low SSR environment, micro-carriers would be designed to collapse when local SSR is below the given threshold.

$$k_c = \begin{cases} k_{col} & \text{if } \dot{\gamma} \leq \dot{\gamma}_{threshold} \\ 0 & \text{if } \dot{\gamma} > \dot{\gamma}_{threshold} \end{cases} \tag{8}$$

The values for the parameters above are presented in Table 2. We propose a novel method to calculate the SSR threshold for the diseased aorta. The determination process of the SSR threshold would be given in section 3.2.

Table 2. Parameters of shear targeted drug delivery simulation

| Parameter | Value | Description |
|---|---|---|
| k | 1.38E-23 J/K | Boltzmann's constant |
| T | 310 K | absolute temperature |
| $\mu_p$ | 1.2E-3 Pa·s | plasma viscosity |
| $r_{micro}$ | 1.5E-6 m | radius of micro-carrier |
| $r_{nano}$ | 9E-10 m | radius of nanoparticle |
| $K_{sh}$ | 5E-2 | dimensionless coefficient |
| $r_{RBC}$ | 4E-6 m | radius of red blood cell |
| $k_{col}$ | 1000 s$^{-1}$ | reaction-term constant |
| $\dot{\gamma}_{threshold}$ | 1000 s$^{-1}$ | SSR threshold for coarctation |
| | 50 s$^{-1}$ | SSR threshold for aneurysm |

SSR: shear stress rate.

At the aortic inlet, the concentrations of micro-carriers and nanoparticles were both set to zero. The drug was assumed to target the endothelial cell receptors (Lobatto et al. 2011), zero flux was prescribed at the wall for micro-carriers and nanoparticle drugs. Considering the backflow of the aortic branches, we used the opening boundary condition in the outlet, which allows the blood to leave and enter the aorta. For blood flow out of the aorta, the micro-carriers and nanoparticle drugs were free to flow out. When blood was flowing into the aorta, the previous timestep concentrations of micro-carriers and nanoparticle drugs were specified at the outlets to approximate the backflow concentrations. Table 3 summarizes the boundary conditions for blood, micro-carriers, and nanoparticle drugs.

Table 3. Boundary conditions for blood, micro-carriers and nanoparticle drugs.

| Boundary | blood | micro-carriers | nanoparticle drugs |
|---|---|---|---|
| inlet | parabolic pulsatile waveform | zero | zero |
| wall | rigid and no-slip | zero flux | zero flux |
| outlets | opening and Windkessel | previous timestep | previous timestep |

In the present study, we designed a novel method to inject micro-carriers, which is not only feasible but also can prevent micro-carriers from entering the aortic branches. In Fig. 1, the syringe indicates where the micro-carriers are injected from the near-wall area (1mm from the aortic wall). We performed the injection through source points at a constant release rate (9.5E-6 mol/s), which were averaged by the three injection points in the aortic aneurysm.

## 3. Results

### *3.1 Hemodynamic validation*

The distribution of the blood flow during a cardiac cycle is used to validate our hemodynamic model (Fig. 2). The idealized healthy aorta is the reference of the two diseased aortas. The blood flow crossing the supra-aortic branches accounts for 30.94% of the inflow in the healthy aorta, which agrees well with our assumption and clinical observation (30%). Coarctation of the aorta prevents blood flow to the descending aorta, three aortic branches receive a remarkable increase in blood flow (81.53%). There is no significant change in the blood flow distribution when an aortic aneurysm forms in the descending aorta. Therefore, the blood flow distribution predicted by the present model is accurate and reasonable.

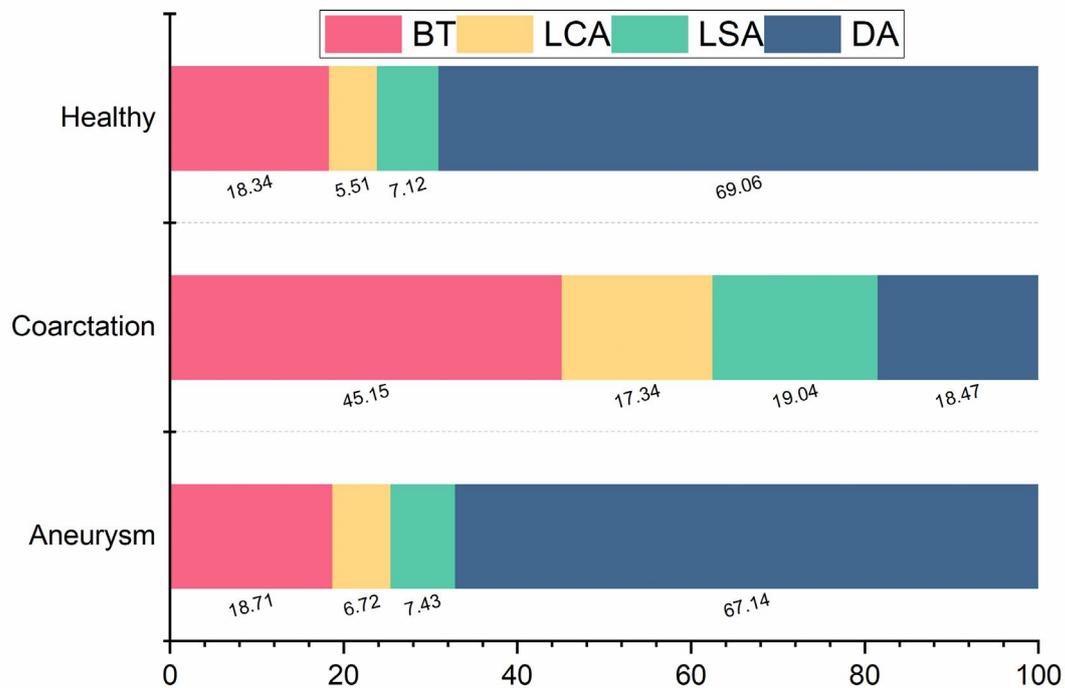

**Fig. 2** Hemodynamic validation. The comparison of the blood flow proportion crossing each outlet among the three cases. The blood flow crossing the supra-aortic branches accounts for 30.94% of the inflow in the healthy aorta, which agrees well with our assumption and clinical observation (30%). (BT: brachiocephalic trunk; LCA: left carotid artery; LSA: left subclavian artery; DA: descending aorta.)

*3.2 Shear stress rate*

Fig. 3 reported the determination process of the SSR thresholds in the present study. For coarctation of the aorta, the surface SSR of the coarctation region and the entire aortic wall is area-averaged during a cardiac cycle. The coarctation region shows a significantly higher SSR value than the entire aortic wall and the SSR threshold is set to 1000 $s^{-1}$ (greater than 975 $s^{-1}$). Unlike the coarctation scenario, the body SSR inside the aneurysm and the entire aorta is volume-averaged during a cardiac cycle. The aneurysm body shows a relatively lower SSR value than the entire aorta body and the SSR threshold is set to 50 $s^{-1}$ (less than 57 $s^{-1}$).

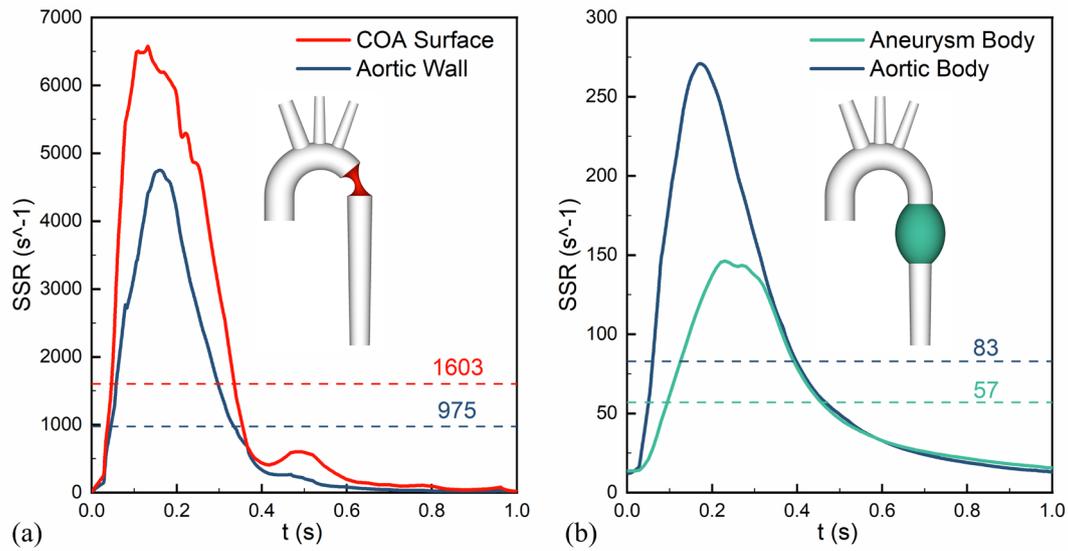

**Fig. 3** Determination process of shear stress rate (SSR) thresholds. (a) Coarctation of the aorta (COA). The surface SSR of the entire aortic wall and coarctation region is area-averaged during a cardiac cycle. The coarctation region shows a significantly higher SSR value than the aortic wall and the SSR threshold is set to 1000 s$^{-1}$ (greater than 975 s$^{-1}$). (b) Aortic aneurysm. The body SSR inside the entire aorta and aneurysm is volume-averaged during a cardiac cycle. The aneurysm body shows a relatively low SSRer value than the aorta body and the SSR threshold is set to 50 s$^{-1}$ (less than 57 s$^{-1}$).

According to the principle of the shear-activated targeted nanoparticle drug delivery system, local abnormal SSR would trigger the micro-carriers to collapse and release nanoparticle drugs. Fig. 4 shows the SSR distribution on the aortic wall and axial cross-section. It should be emphasized that the SSR value was time-averaged (TASSR) over a cardiac cycle. For coarctation of the aorta, the highest TASSR can be seen in the most severe coarctation and brachiocephalic trunk, which provides an ideal SSR environment to activate the micro-carriers. The greater curvature of the aortic arch and proximal descending aorta also expose to high TASSR relative to the lesser curvature of the aortic arch and distal descending aorta. However, the interior of the aorta is in a low-TASSR environment (Fig. 4b). It is worth noting that the maximum SSR value of the legend is the threshold value we adopt. In the aortic aneurysm, the area of interest is the enlarged aneurysm wall, which shows the lowest TASSR. The interior of the aortic aneurysm is also in a low shear rate environment, which is used to activate the micro-carriers. The time-averaged velocity is also illustrated in Fig. 4. A high-speed jet is observed in the coarctation and flow separation occurs after the most severe coarctation area, which coincides with the high SSR region. For the aortic aneurysm, the presence of backflow results in a low-velocity area, which exposes to low SSR value.

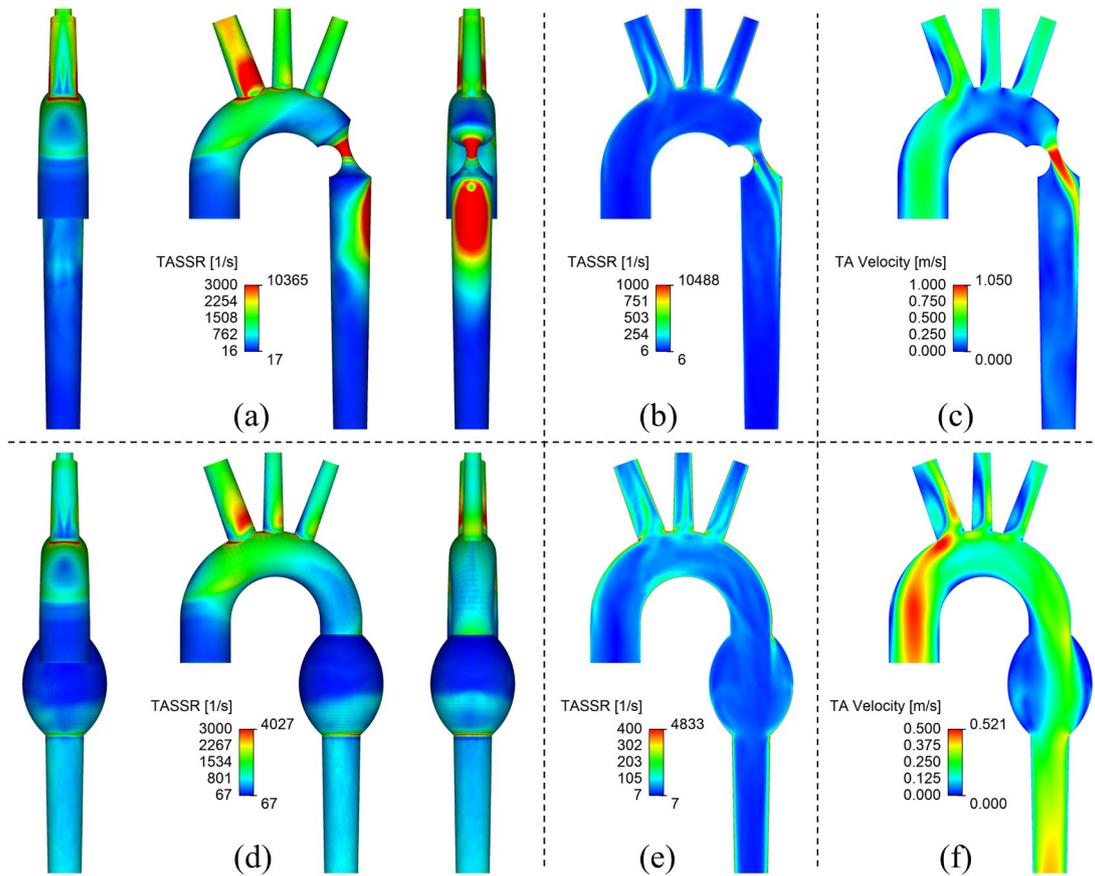

**Fig. 4** Distribution of time-averaged shear stress rate (TASSR) and velocity. Coarctation of the aorta: (a) Three views of the wall, (b) Axial cross-section, and (c) Velocity in cross-section. Aortic aneurysm: (d) Three views of the wall, (e) Axial cross-section, and (f) Velocity in cross-section. The extreme values are shown on the right side of the color axes.

### *3.3 Micro-carriers and nanoparticle drugs*

Fig. 5 investigates the time-averaged concentration of micro-carriers and nanoparticle drugs on the aortic wall. The accumulated concentration of released nanoparticles on the diseased aortic wall can characterize the efficiency of shear-activated targeted drug delivery. For coarctation of the aorta, most micro-carriers are activated in the most severe coarctation, which exhibits a high concentration of nanoparticle drugs along with the descending aorta downstream. It should be noted that a fewer nanoparticle drug released in advance by relatively high SSR of the injection position could reach the proximal wall of the coarctation compared to the distal coarctation and descending aorta. In the aortic aneurysm, almost no micro-carriers accumulate on the aortic wall, which indicates that most micro-carriers successfully transform into nanoparticle drugs when flowing through the aneurysm. Moreover, it can be observed that the released nanoparticle drugs accumulate on the aneurysm wall.

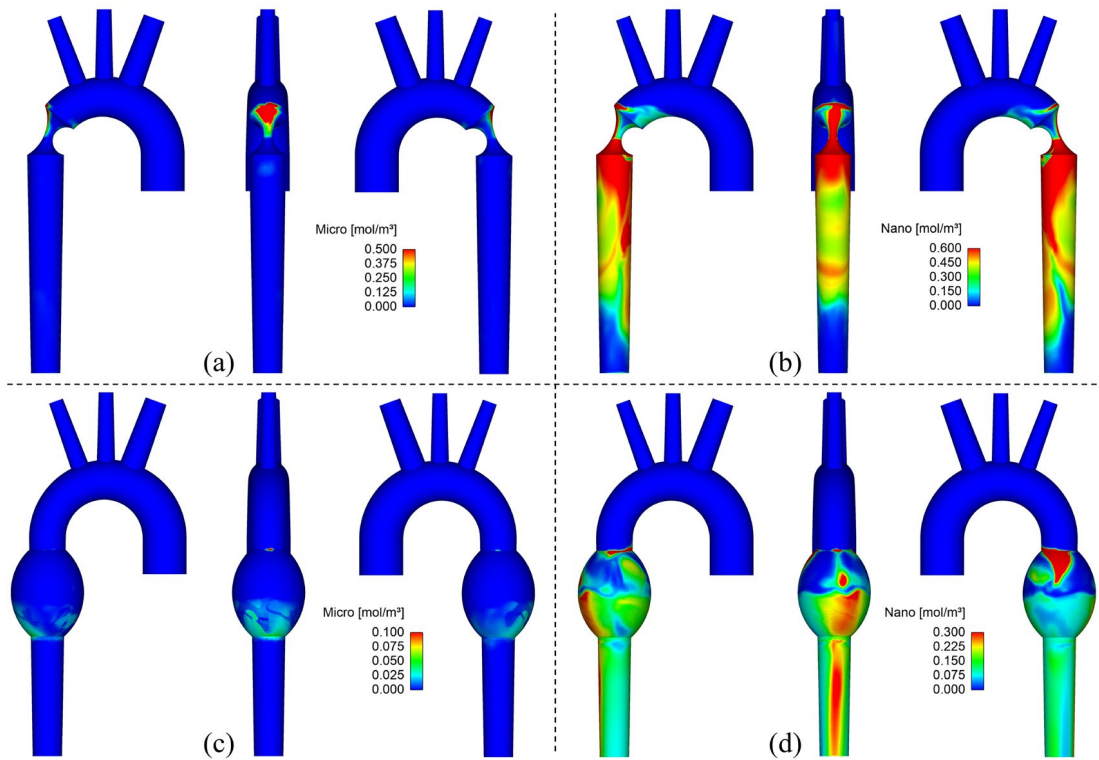

**Fig. 5** Time-averaged concentrations of micro-carriers and nanoparticle drugs on the aortic wall. Coarctation of the aorta: (a) micro-carriers and (b) nanoparticle drugs. Aortic aneurysm: (c) micro-carriers and (d) nanoparticle drugs.

The time-averaged concentrations of micro-carriers and nanoparticle drugs in the interior of the aorta are depicted in Fig. 6. For coarctation of the aorta, micro-carriers accumulate before the coarctation and part of it enters the descending aorta through coarctation. The presence of micro-carriers in the descending aorta indicates that the SSR threshold could not release all the nanoparticle drugs. In contrast to coarctation of the aorta, the low-shear environment of the aortic aneurysm activates all the micro-carriers. Nanoparticle drugs fill the entire aneurysm before leaving the low SSR region.

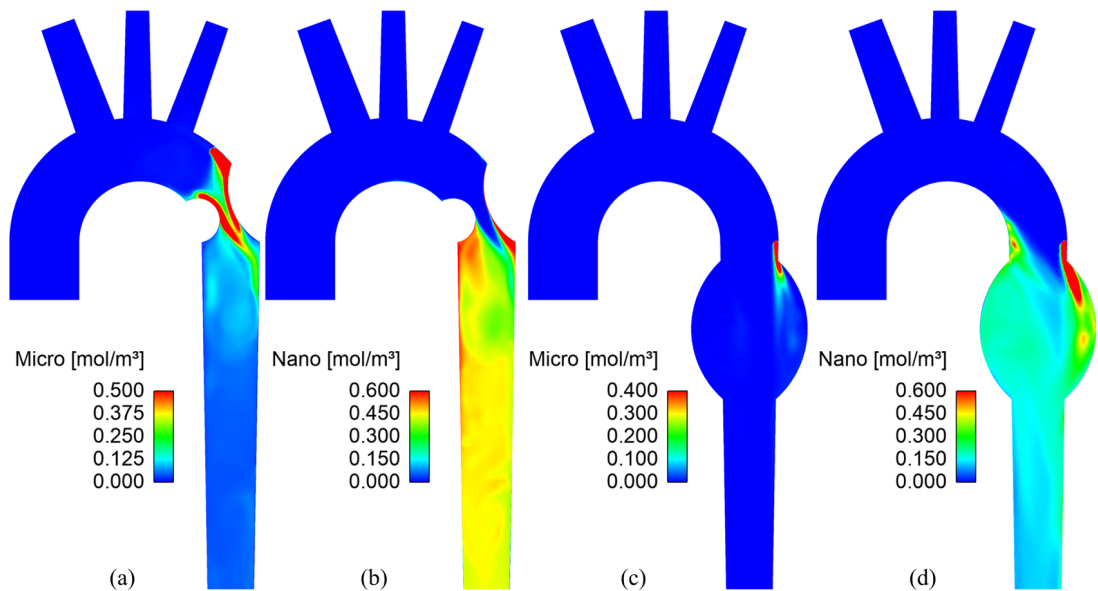

**Fig. 6** Time-averaged concentrations of micro-carriers and nanoparticle drugs on the axial cross-sections. Coarctation of the aorta: (a) micro-carriers and (b) nanoparticle drugs. Aortic aneurysm: (c) micro-carriers and (d) nanoparticle drugs.

*3.4 Diffusivity of the micro-carriers and nanoparticle drugs*

Fig. 7 illustrates the impact of the high diffusivity of the micro-carriers and nanoparticle drugs on the surface drug concentrations. The diffusion coefficients are expanded by 100 times. For coarctation of the aorta, the concentrations of the micro-carriers and nanoparticle drugs are pulsating and high diffusivity enhances the oscillation amplitude of the drug concentration. Eventually, the concentration of micro-carriers increases while the concentration of nanoparticle drugs decreases. In the aortic aneurysm, only the micro-carriers concentration shows pulsating characteristics while the magnitude of nanoparticle drugs is continuously rising. It should be emphasized that the final surface concentration of nanoparticle drugs is remarkably elevated.

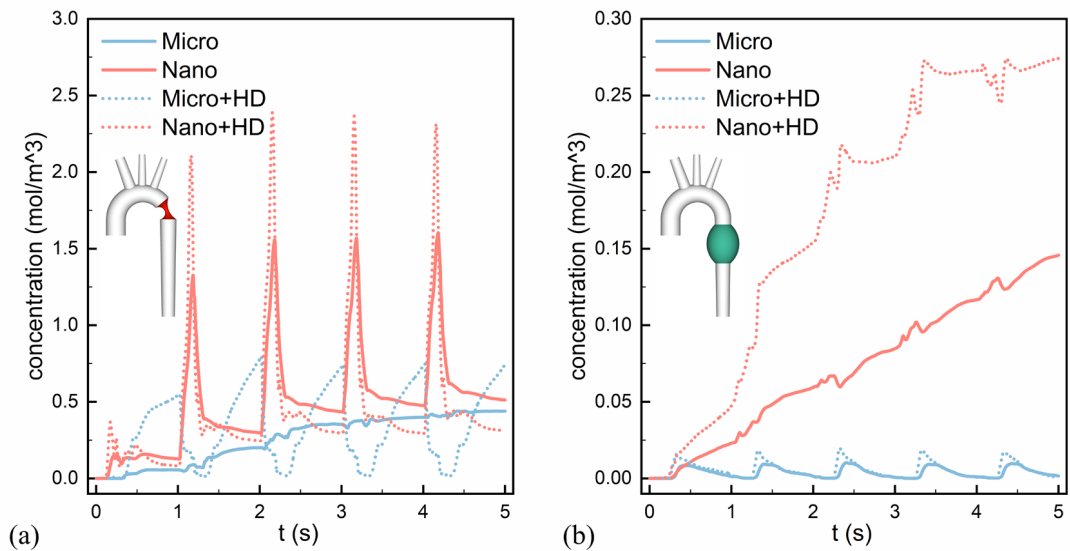

**Fig. 7** Effect of the high diffusivity of the micro-carriers and nanoparticle drugs on the surface drug concentrations. (a) Coarctation of the aorta. (b) Aortic aneurysm. HD indicates high diffusivity.

## 4. Discussion

The clinical treatment of aortic diseases is developing towards minimally invasive and high survival rates. Targeted drug delivery therapy has remarkable potential to fulfill this ambitious goal. If the pathogenesis of aortic diseases is revealed, specific drugs will be manufactured. This primary study aims to evaluate the feasibility of shear-activated targeted nanoparticle drug delivery in the treatment of coarctation of the aorta and aortic aneurysm by using computational fluid dynamics.

The choice of micro-carriers injection position is very important, which could be performed by smart micro-needles and micro-catheters (Kleinstreuer et al. 2014; Tibbitt et al. 2016). For coarctation of the aorta, there is an abnormal high SSR hemodynamic environment when blood flows through the most severe coarctation. Therefore, the micro-carrier should be designed to be stable under low SSR. Based on this condition, we first chose the center point of a radial section between the left subclavian artery and coarctation, where the SSR was low enough and the injection position could prevent micro-carriers from entering the aortic branches. However, we found the micro-carriers could only be activated at the most severe coarctation and there were no nanoparticle drugs in proximal coarctation. Therefore, the injection position was moved to the near-wall area before the coarctation, where the SSR was relatively high and part of micro-carriers would be activated in advance. The situation with an aortic aneurysm was just the opposite. A low SSR environment was observed in the aortic aneurysm, micro-carriers were designed to collapse when the SSR was abnormally reduced. We chose the upstream near-wall area with high SSR to inject micro-carriers, which can avoid premature activation of micro-carriers. When flowing into the aneurysm body, the micro-carriers would be activated by the abnormal body SSR. Therefore, the upstream near-wall area with high SSR is an idealized position to inject micro-carriers.

Meschi et al. (2021) has reported the release process of nanoparticle drugs in coronary artery disease and a 2D plane was used to display the simulation results. As mentioned above, the concentration of nanoparticles on the diseased aortic wall could be used to describe the efficiency of shear-activated targeted drug delivery. However, we found that the distribution of the nanoparticle drugs on the aortic wall is different from interior distribution, indicating that the value of a cross-section cannot represent the aortic external wall distribution. In summary, the concentration distribution of nanoparticle drugs on the blood vessel wall is suggested to present in targeted nanoparticle drug delivery study.

We found that the high diffusivity of the micro-carriers and nanoparticle drugs has a significant impact on the surface drug concentrations of the diseased aortic walls. For coarctation of the aorta, the diffusivity of the drug should be controlled in a reasonable range to ensure enough drugs reach the coarctation wall. While aneurysm shows different characteristics and high diffusivity would increase the final surface concentration of nanoparticle drugs. Considering the enlarged volume of the aortic aneurysm, the diffusivity of the micro-carriers and nanoparticle drugs needs to be improved for high efficiency.

In this pilot study, the strategy of shear-activated targeted nanoparticle drugs was firstly applied in the treatment of coarctation of the aorta and aortic aneurysm. The feasibility of this new strategy of targeted drug delivery for treating aortic diseases was evaluated. There are still some limitations that need to be emphasized. First, the aortic wall was assumed to be rigid and the interaction between the blood flow and aortic wall was neglected. Our previous study has shown that fluid-structure interaction has a slight effect on the distribution of wall shear stress in simple aortic geometry such as aneurysm (Qiao et al. 2021; Qiao et al. 2019b). Additionally, the core hemodynamic parameter in the present mathematic model of shear-activated targeted nanoparticle drugs is SSR. Therefore, the assumption of a rigid wall is a reasonable compromise with the limited computational resource. Second, the aortic geometric model is idealized and patient-specific cases of aortic diseases would be explored in our future study. Finally, specific nanoparticle drugs for the treatment of aortic diseases are still under research. Validation with experimental data would be carried out in future studies.

## 5. Conclusions

This study evaluates the feasibility of shear-activated targeted nanoparticle drug delivery in aortic diseases by using computational fluid dynamics. We propose novel methods to inject the micro-carriers and calculate the SSR thresholds for coarctation of the aorta and aortic aneurysm. Our results demonstrate that this promising strategy can be applied in the treatment of coarctation of the aorta and aortic aneurysm. The micro-carriers injected from the upstream near-wall of the diseased location could be activated by the abnormal SSR. Released nanoparticle drugs would be successfully targeted delivered to the aortic diseased wall. Besides, the high diffusivity of the micro-carriers and nanoparticle drugs has a remarkable impact on the surface drug concentration. The unique hemodynamic environment of diseased blood vessels could be used in the design process of a more individualized and efficient novel treatment.